\documentclass[pra,aps,twocolumn,floatfix,showpacs]{revtex4}
\usepackage{graphicx,amsmath,amssymb,times}

\begin{document}
\title{Dynamical mean-field theory for light fermion--heavy boson mixtures on optical lattices}
\author{M. Iskin$^1$ and J. K. Freericks$^2$}
\affiliation{$^1$ Department of Physics, Ko\c c University, Rumelifeneri Yolu, 34450 Sariyer, Istanbul, Turkey. \\
$^2$Department of Physics, Georgetown University, Washington, DC 20057, USA.}
\date{\today}

\begin{abstract}
We theoretically analyze Fermi-Bose mixtures consisting of light fermions
and heavy bosons that are loaded into optical lattices 
(ignoring the trapping potential). To describe such mixtures, 
we consider the Fermi-Bose version of the Falicov-Kimball model on a periodic lattice. 
This model can be exactly mapped onto the spinless Fermi-Fermi Falicov-Kimball model 
at zero temperature for all parameter space as long as the mixture is thermodynamically stable. 
We employ dynamical mean-field theory to investigate the evolution of the 
Fermi-Bose Falicov-Kimball model at higher temperatures. We calculate spectral 
moment sum rules for the retarded Green's function and self-energy, 
and use them to benchmark the accuracy of our numerical calculations, as well as 
to reduce the computational cost by exactly including the tails of infinite 
summations or products. We show how the occupancy of the bosons, single-particle 
many-body density of states for the fermions, momentum distribution, and the 
average kinetic energy evolve with temperature. 
We end by briefly discussing how to experimentally realize 
the Fermi-Bose Falicov-Kimball model in ultracold atomic systems.
\end{abstract}

\pacs{03.75.Lm, 37.10.Jk, 67.85.-d, 67.85.Pq}
\maketitle

\section{Introduction}
\label{sec:introduction}
Experimental work in ultracold atomic systems in optical lattices has 
been progressing rapidly. First generation experiments focused primarily 
on single species systems, or mixtures of different isotopes of the same 
atomic species. Now, experimental work is progressing into the realm of 
mixtures of different species of atoms. Since these atomic species often 
have significantly different masses, one immediately expects such systems 
to respond differently than isotopic mixtures. In addition, if light alkali 
atoms like Li or K are used, or if one uses alkaline earth atoms like 
Sr or rare-earths like Yb, then one has the prospect for modifying 
the particle statistics from Fermi-Dirac to Bose by simply changing the 
isotope employed in the experiment. 

In this work, we focus on Fermi-Bose mixtures where the bosonic atoms 
are the heavy atoms (which is typically the experimental situation if Li 
or K is the light fermion and Rb, Cs, Sr, or Yb is the heavy boson).  
While much work on Fermi-Bose mixtures has focused on either how the 
presence of the fermions modifies the Bose-Einstein condensation (BEC) 
of the bosons, or how the presence of the bosons modifies the interactions 
of the fermions (\textit{e.g.} by allowing phonons)~\cite{bose_fermi_mod}, 
our work here focuses on the opposite situation where the heavy bosons 
are so heavy that we can ignore the quantum-mechanical effects of their 
kinetic energy, and hence they never condense into a BEC. Instead, at low 
temperature, the mixture either phase separates or forms static density 
wave patterns, which can be quite complex. We believe such a situation 
arises when the tunneling amplitude of the heavy bosons on 
the optical lattice is more than an order of magnitude smaller 
than the tunneling amplitude of the light fermions.  
This typically occurs when the lattice depth is deep (more than 10 to 
15 recoil energies of the Rb for a K-Rb mixture) and the system is 
well represented by a single-band model~\cite{prl_dipolar_molecules}.  
In this situation, experiments are likely to be run at temperatures 
significantly higher than the BEC temperature, and it is more likely 
that static density waves would form instead of superfluidity 
(in other words, we are examining how the fermions modify the 
Mott-insulating state of the bosons, which occurs when the boson-boson 
repulsion is much larger than the boson tunneling amplitude~\cite{fisher_etal}). 
Long-range effective boson-boson interactions are generated via the 
interactions with the mobile fermions. This realm is not so well 
known within the atomic physics community, although similar models have 
been widely studied within the condensed matter physics community, 
as we describe below.

Recently, mixtures of fermionic $^{40}$K and bosonic $^{87}$Rb atoms have
been first studied at fixed interspecies interaction strengths by two 
different experimental groups~\cite{gunter, ospelkaus}, and later with tunable 
interactions~\cite{best}, where a shift of the bosonic superfluid to 
Mott insulator transition has been observed due to the interspecies 
coupling (irrespective of its sign but with significant asymmetry).
There are several theoretical proposals to explain this 
effect~\cite{cramer, luhmann, pollet, refael, lutchyn}.
Motivated by these experiments, here we study such light-Fermi--heavy-Bose 
mixtures with the Fermi-Bose version of the Falicov-Kimball (FK) 
model~\cite{falicov_kimball,ates_ziegler,vollhardt},
the Fermi-Fermi version of which has been widely discussed 
in the condensed matter literature~\cite{gruber_macris,jkf_review}. 
In our case, the bosons have no quantum dynamics of motion, but they 
can sample all possible heavy-atom configurations in an annealed 
statistical-mechanical sense.

Our main results are as follows.
First, we examine the symmetries of the Hamiltonian, and show that the 
Fermi-Bose FK model can be mapped exactly onto the spinless Fermi-Fermi 
FK model at zero temperature for all parameter space as long as the 
mixture is thermodynamically stable. Since this mapping is only approximate at low
temperatures and it fails at high temperatures, we develop dynamical 
mean-field theory (DMFT)~\cite{brandt_mielsch} (which becomes exact in 
infinite dimensions) to investigate the effects of temperature 
and how the Fermi-Bose system evolves into an effective Fermi-Fermi system.
In addition, we calculate spectral moment sum rules for the retarded 
Green's function and self-energy, and use them to check the accuracy of our 
numerical calculations, as well as to reduce the computational cost~\cite{hubb_sum,sum1,sum2,sum3}. 
We also present typical numerical results for the Fermi-Bose FK model 
including the occupancy of the bosons, single-particle many-body
density of states (DOS) for the fermions, momentum distribution, and
the average kinetic energy. 
The DOS has significant temperature dependence as the system evolves 
from high to low temperatures. For example, in the insulating regime, 
where the magnitude of the boson-fermion interaction is much larger 
than the magnitude of the fermion tunneling amplitude, many `upper Hubbard bands' 
exist at high temperature, with the weights of each band determined 
by the probability that the bosons multiply occupy a lattice site. 
These bands then evolve to just one upper and one lower Hubbard band at low temperature.

The remainder of this manuscript is organized as follows. In Sec.~\ref{sec:mixtures}, 
we first introduce the Hamiltonian for the Fermi-Bose version of the FK model,
and then discuss its symmetries. In Sec.~\ref{sec:dmft}, we develop 
the DMFT for this model, where we also discuss spectral moment sum rules for 
the retarded Green's function and self-energy. The numerical results 
obtained from the DMFT are discussed in Sec.~\ref{sec:numerics}, 
and a brief summary and conclusions are presented in Sec.~\ref{sec:conclusions}.

\section{Light Fermion--Heavy Boson Mixtures}
\label{sec:mixtures}

We analyze Fermi-Bose mixtures consisting of light fermions and 
heavy bosons that are loaded into optical lattices. We consider the 
limiting case such that the band mass of bosons 
$M_b$ is much greater than that of fermions ($M_b \gg M_f$). In this case, 
the tunneling term for the bosons $t_b$ is much smaller than that of the 
fermions ($t_b \ll t_f$), and we may set $t_b = 0$ in the quantum-mechanical 
Hamiltonian that describes the system. Such mixtures could 
be experimentally realized in ultracold atomic systems. 
For instance, a $^{40}$K-$^{87}$Rb mixture with $M_b\approx 2.2M_f$, 
a $^{6}$Li-$^{41}$K mixture with $M_b \approx 6.8 M_f$, 
or a $^{6}$Li-$^{133}$Cs mixture with $M_b \approx 22.2 M_f$ are 
good candidates for the applicability of this Hamiltonian. In addition, one could also
create species-dependent optical lattices for different isotopes of the same 
atom such that the bosonic isotope is localized but the fermionic one is not, 
which might be most relevant for isotopic mixtures of Sr or Yb 
(in other words, it isn't the masses {\it per se} that need to be different, 
but it is the tunneling amplitudes when in an optical lattice that need 
to be quite different). It is important to emphasize that, although we consider the limit of 
localized bosons, the problem at hand is still a strongly correlated 
many-body problem since the bosons and fermions are coupled by the 
interaction and we take annealed statistical averages over all heavy-atom configurations.

\subsection{The Fermi-Bose Falicov-Kimball (FK) Model}
\label{sec:fc}

With such systems in mind, we analyze the following Hamiltonian to 
describe the light fermion--heavy boson mixtures~\cite{falicov_kimball,ates_ziegler,vollhardt}
\begin{align}
\label{eqn:hamiltonian}
H &= -t_f\sum_{\langle i,j\rangle} (f_i^\dagger f^{}_j + f_j^\dagger f^{}_i) + 
\frac{U_{bb}}{2} \sum_i b_i^\dagger b^{}_i (b_i^\dagger b^{}_i - 1)  \nonumber \\
&+ U_{bf} \sum_i b_i^\dagger b^{}_i f_i^\dagger f^{}_i - \mu_f \sum_i f_i^\dagger f^{}_i - \mu_b \sum_i b_i^\dagger b^{}_i,
\end{align}
where $f_i^\dagger$ ($f_i$) is the creation (annihilation) operator for 
an itinerant spinless fermion at site $i$, and $b_i^\dagger$ ($b_i$) is the corresponding
operator for a localized boson at site $i$. The fermionic operators satisfy 
the usual canonical anticommutation relations $\{f_i,f_j^\dagger\} = \delta_{ij}$
where $\delta_{ij}$ is the Kronecker delta function, 
the bosonic operators satisfy the usual canonical commutation relations 
$[b_i,b_j^\dagger] = \delta_{ij}$, and the fermionic and bosonic 
operators all mutually commute. The first term is the kinetic
energy of the fermions with $t_f$ denoting the nearest-neighbor tunneling. 
The second and third terms are the on-site density-density interactions 
between the bosons themselves, and between the bosons and fermions, respectively. 
The on-site fermion-fermion interaction is not allowed because of 
the Pauli exclusion principle, \textit{i.e.} $U_{ff} \to \infty$.
The last two terms involve the chemical potentials of the fermions
($\mu_f$) and bosons ($\mu_b$), that are employed to adjust the 
filling of the corresponding particles to the desired values.

The model Hamiltonian given in Eq.~(\ref{eqn:hamiltonian}) then 
corresponds to the Fermi-Bose version of the FK model, the 
Fermi-Fermi version of which has been studied extensively in the 
condensed-matter literature~\cite{gruber_macris,jkf_review}. 
At low temperatures, the Fermi-Fermi version of the model is known 
to possess significant regions of density wave order with complicated 
patterns~\cite{gruber_macris}, and also has a strong tendency to phase 
separate when the light fermion--heavy fermion interaction is 
large and repulsive~\cite{lieb}.  
It is also known that the heavy fermions cannot generate an effective 
retarded light fermion--light fermion attraction leading to 
superconductivity as a consequence of Anderson's theorem~\cite{anderson}.  
Within the DMFT context, essentially all response functions 
(both static and dynamical) and all kinds of transport have been 
evaluated for the system~\cite{jkf_review}. We show below that 
many of these low temperature properties are shared by the Fermi-Bose 
FK model, but at higher temperatures, the behavior is quite different.

\subsection{Symmetries of the Hamiltonian}
\label{sec:symmetry}

This Hamiltonian possesses partial particle-hole symmetries for the
fermions and bosons. These symmetries hold on a bibartite lattice 
where it is possible to divide the entire lattice into two sublattices 
$A$ and $B$ such that the fermionic tunneling only connects 
different sublattices. When the particle-hole transformation
is applied to the fermions, \textit{i.e.} $f_i^\dagger \to f_i^h (-1)^{p(i)}$ 
and $f_i \to (f_i^h)^\dagger (-1)^{p(i)}$ with $p(i) = 0$ for $i \in A$
and $p(i) = 1$ for $i \in B$, then, up to a numerical shift, it can be 
shown that the Hamiltonian that is expressed in terms of the 
hole operators for the fermions maps onto the starting 
Hamiltonian with $\mu_f \to -\mu_f$, $U_{bf} \to -U_{bf}$ 
and $\mu_b \to \mu_b - U_{bf}$. In the canonical ensemble, 
this means that the energies are simply related by
$
E(\rho_f, \rho_b, U_{bf}) \equiv E(1-\rho_f, \rho_b, -U_{bf}),
$
where $\rho_f$ and $\rho_b$ are the fillings for fermions and bosons,
respectively. We notice that this mapping for the fermions holds 
at any temperature $T$, unlike that for the bosons,  discussed next.

When the particle-hole transformation is applied to the bosons, 
\textit{i.e.} $b_i^\dagger \to b_i^h (-1)^{p(i)}$ and $b_i \to (b_i^h)^\dagger (-1)^{p(i)}$,
then, up to a numerical shift, it can also be shown that the 
Hamiltonian that is expressed in terms of the hole operators for 
the bosons maps onto the starting Hamiltonian with
$\mu_f \to \mu_f + U_{bf}$ and $\mu_b \to \mu_b + U_{bb}$.
In the canonical ensemble, this means that the energies of many-body eigenstates
are simply related by
$E(\rho_f, \rho_b, U_{bf}) \equiv E(\rho_f, 1+\rho_b, U_{bf})$.
This is not enough to map Green's functions onto each other.  The subtle issue is that the
particle states with $n_b = 0, 1, 2, \cdots, \infty$ 
map onto the hole states $n_b^h \equiv n_b + 1 = 1, 2, 3, \cdots, \infty$, hence we
lose all information about the weight of the $n_b^h = 0$ state when more than one $n_b>0$ state 
has a nonzero density; when only two states are nonzero, the mapping above plus 
the relation between particle and hole densities allows us to find all of the 
bosonic weights. This occurs exactly only at $T=0$ where at most two states 
have nonzero weights. Hence the bosonic particle-hole symmetry 
only occurs at $T=0$. This symmetry also holds in the 
atomic ($t_b = 0$) limit of the Bose-Hubbard model at $T = 0$, 
where the energy for the particle excitations of the $n_b$th 
state is degenerate with that for the hole excitations of 
the ($n_b + 1$)th state, but it does not hold when $t_b\ne 0$, because 
one can no longer label the many-body eigenstates by the local boson particle number.

The latter symmetry implies that there can be at most two bosonic 
states that can be occupied at $T = 0$. For instance, 
$n_b = \{0,1\}$ states for $0 < \rho_b < 1$; 
$n_b = 1$ state for $\rho_b = 1$; 
$n_b = \{1,2\}$ states for $1 < \rho_b < 2$; 
$n_b = 2$ state for $\rho_b = 2$; 
$n_b = \{2,3\}$ states for $2 < \rho_b < 3$; etc. 
This is similar to what happens in the spinless Fermi-Fermi FK model 
where there can be at most two fermionic states for the heavy 
fermions due to the Pauli exclusion principle. Therefore, the Fermi-Bose 
FK model can be exactly mapped onto the well-studied spinless Fermi-Fermi 
FK model for all parameter space at $T = 0$ as long as the mixture is thermodynamically stable. 
Since this mapping is only approximate at low temperatures, 
and it fails at high-temperatures, we develop DMFT to investigate the effects of temperature.

\section{Dynamical mean-field theory}
\label{sec:dmft}

The DMFT for the Fermi-Fermi FK model is well established in the 
literature~\cite{brandt_mielsch,jkf_review}, and it can be easily 
generalized to the Fermi-Bose case that we consider here. For this
purpose, we notice that the bosons cannot undergo BEC since there 
are a fixed number of localized bosons on each lattice site and 
the local boson particle number is an operator that commutes 
with the Hamiltonian. Then the impurity partition function can be 
integrated analytically since the effective action is quadratic 
in the fermionic operators and depends only on the local bosonic 
number operators. As a result, similar to the Fermi-Fermi case, there are 
three main equations that need to be solved self-consistently for the 
Fermi-Bose FK model.

\subsection{Imaginary-axis Formalism}
\label{sec:iaxis}

The first equation is the single-particle lattice Green's function for
the fermions defined by the time-ordered product
$
G(\tau) = -\langle T_\tau f_i(\tau) f_i^\dagger(0)  \rangle
$
in the imaginary time ($0 \le \tau \le \beta$) representation, where $T_\tau$ 
is the imaginary time ordering operator, and
$
\langle O \rangle \equiv \mathrm{Tr} \{e^{-\beta H} O\}/Z
$
is the ensemble average with $Z = \mathrm{Tr} \{e^{-\beta H}\}$ the 
partition function and $\beta = 1/T$ the inverse temperature. Here, the 
creation and annihilation operators are in the imaginary-time Heisenberg representation
$
O(\tau) = e^{H\tau} O(0) e^{-H\tau}.
$
In the Matsubara frequency representation,
$
G(i\omega_n) = \int_{0}^{\beta} d\tau e^{i\omega_n \tau} G(\tau),
$
where $\omega_n = (2n+1) \pi/\beta$ is the fermionic Matsubara frequency, 
the Green's function becomes
\begin{equation}
\label{eqn:G}
G(i\omega_n) = \sum_{n_b = 0}^{\infty} \frac{w_{n_b}}{G_0^{-1}(i\omega_n)- U_{bf}n_b},
\end{equation}
where $w_{n_b} = Z_{n_b}/Z$ (with $Z = \sum_{n_b = 0}^{\infty} Z_{n_b}$) is the 
probability for each site to be occupied exactly by $n_b = 0, 1, 2, \cdots, \infty$ bosons,
and $G_0^{-1}(i\omega_n) = i\omega_n + \mu_f - \lambda(i\omega_n)$
is the bare Green's function with $\lambda(i\omega_n)$ the dynamical mean field.
Here,
\begin{equation}
\label{eqn:Znb}
Z_{n_b} =  e^{\beta[\mu_b n_b - U_{bb}n_b(n_b-1)/2]} Z_0(\mu_f-U_{bf} n_b)
\end{equation}
is the partition function for the $n_b$ state, where
\begin{equation}
\label{eqn:Z0}
Z_0(\mu_f) = 2e^{\beta \mu_f/2} \prod_{n = -\infty}^{\infty} \frac{i\omega_n + \mu_f - \lambda(i\omega_n)}{i\omega_n}
\end{equation}
is the fermionic partition function for $U_{bf} = 0$. The prefactor in Eq.~(\ref{eqn:Z0}) 
is added to give the correct noninteracting result when $\lambda(i\omega_n) = 0$. 
In the case of the spinless Fermi-Fermi FK model, due to the Pauli exclusion 
principle, \textit{i.e.} the heavy fermion--heavy fermion interaction $U_{bb} \to \infty$, 
only two states ($w_0$ and $w_1$) can be occupied such that $w_1 = 1-w_0$
is the filling of localized fermions. Note these equations are similar to 
the solution of the classical Holstein model~\cite{millis}, but here the 
boson states are discrete, while there they are continuous.

The partition functions also satisfy the well-known relations 
$Z_0(\mu_f) = \mathrm{Det} G_0^{-1}(i\omega_n)$ and $Z = \mathrm{Det} G^{-1}(i\omega_n)$, 
so that the bare Green's function can be re-expressed as
\begin{align}
\label{eqn:G0}
G_0^{-1}(i\omega_n) = G^{-1}(i\omega_n) + \Sigma^{-1}(i\omega_n),
\end{align}
which is our second equation. This is Dyson's equation which relates the 
bare Green's function to the self-energy $\Sigma(i\omega_n)$, and can 
also be thought as the definition of the self-energy.

The third equation is given by the lattice Hilbert transform of the noninteracting DOS,
\begin{equation}
\label{eqn:Gk}
G(i\omega_n) = \int_{-\infty}^{\infty} \frac{d\epsilon \rho(\epsilon)}{i\omega_n + \mu_f - \Sigma(i\omega_n) - \epsilon},
\end{equation}
where
$
G(i\omega_n) = \sum_{\mathbf{k}} G(\mathbf{k}, i\omega_n)
$
with 
$
G(\mathbf{k}, i\omega_n) = 1/[i\omega_n + \mu_f - \Sigma(i\omega_n) - \epsilon(\mathbf{k})]
$
the momentum-resolved Green's function, and
$
\rho(\epsilon) = \sum_{\mathbf{k}} \delta[\epsilon - \epsilon(\mathbf{k})] 
= e^{-(\epsilon/t^*)^2}/(\sqrt{\pi} t^*)
$ 
is the noninteracting DOS for the infinite-dimensional hypercubic lattice with 
$\delta(x)$ the delta function and $\epsilon(\mathbf{k}) = -2t_f\sum_{i = 1}^{d} \cos(k_i a)$ 
the energy dispersion for the fermions. Here, we used the fact that the tunneling
$t_f$ scales with dimension $d$ such that $t_f = t^*/\sqrt{4d}$~\cite{metzner_vollhardt}.
This equation can be rewritten in terms of the Faddeeva function
$
G(i\omega_n) = -i\sqrt{\pi} e^{-[i\omega_n + \mu_f - \Sigma(i\omega_n)]^2} 
\mathrm{erfc}\{-i \eta [i\omega_n + \mu_f - \Sigma(i\omega_n)]\},
$
where 
$
\mathrm{erfc}(z) = (2/\sqrt{\pi}) \int_z^{\infty} dz e^{-z^2}
$
is the complex complementary error function, and 
$\eta = \mathrm{Sign}\{\mathrm{Im}[i\omega_n - \Sigma(i\omega_n)]\}$.

It is convenient to solve Eqs.~(\ref{eqn:G}),~(\ref{eqn:G0}) 
and~(\ref{eqn:Gk}) self-consistently by using an iterative approach~\cite{jarrell}.
For a fixed set of $U_{bb}$, $U_{bf}$ and $T$, our strategy 
is as follows: 
(1) choose initial values for $\mu_f$ and $\mu_b$;
(2) start with an initial value for $\Sigma(i\omega_n)$, \textit{e.g.} $\Sigma(i\omega_n) = 0$, 
and plug it into Eq.~(\ref{eqn:Gk}) to solve for $G(i\omega_n)$;
(3) use $G(i\omega_n)$ and the initial $\Sigma(i\omega_n)$ in Eq.~(\ref{eqn:G0})
to obtain $G_0(i\omega_n)$;
(4) plug $G_0(i\omega_n)$ into Eq.~(\ref{eqn:G}) to solve for the new $G(i\omega_n)$;
(5) use $G_0(i\omega_n)$ and the new $G(i\omega_n)$ in 
Eq.~(\ref{eqn:G0}) to obtain a new $\Sigma(i\omega_n)$; and
(6) iterate steps (2)--(5) with the new $\Sigma(i\omega_n)$ until the solution 
converges.
Finally, we (7) adjust $\mu_f$ and $\mu_b$ until the desired 
$\rho_f = \langle f_i^\dagger f_i \rangle$ and $\rho_b = \langle b_i^\dagger b_i \rangle$ 
values are reached. 
Typically, the self-energy converges to eight decimal points in less than 
twenty iterations (in a few seconds) for each choice of $\mu_f$ and $\mu_b$,
and the entire procedure takes less than a minute to obtain the particular
values for $\rho_f$ and $\rho_b$.

\subsection{Real-axis Formalism}
\label{sec:raxis}

Once the chemical potentials $\mu_f$ and $\mu_b$, and the occupation probabilities 
$w_{n_b}$ are calculated from the imaginary-axis calculation,  we can employ the 
analytic continuation of Eqs.~(\ref{eqn:G}),~(\ref{eqn:G0}) and~(\ref{eqn:Gk}) 
with $i\omega_n \to \omega + i0^+$, to calculate $G(\omega)$ 
and $\Sigma(\omega)$ on the real axis using a similar iterative approach.
Typically, the convergence is slower on the real axis than on the imaginary 
one, especially near the correlation induced band edges which leads to a number
of poles in the self-energy when $U_{bf}$ is large enough. 
Therefore, it is important to verify the consistency between the
imaginary- and real-axis calculations. 

For instance, one of the stringent tests is the comparison of the Green's function
that is calculated directly on the imaginary axis by the algorithm described 
above to the spectral representation for the imaginary axis Green's function given by
$
G(i\omega_n) = -(1/\pi) \mathrm{Im} \int_{-\infty}^{\infty} d\omega G(\omega) / (i\omega_n - \omega)
$, with the Green's function generated by the real-axis code appearing in the integrand.  
We usually achieve more than four digits of accuracy between the two calculations.
Another test is the comparison of the filling of the fermions that can be calculated directly
on the imaginary axis via
$
\rho_f = (1/\beta) \sum_{n=-\infty}^{\infty} G(i\omega_n)
$
to that found from the real axis with
$
\rho_f = -(1/\pi) \int_{-\infty}^{\infty} d\omega F(\omega) \mathrm{Im}G(\omega),
$
where $F(\omega) = 1/(1+e^{\beta \omega})$ is the Fermi-Dirac distribution function.

One can also check the spectral moment sum rules for the retarded Green's 
function, and also for the retarded self-energy~\cite{hubb_sum,sum1,sum2,sum3}. 
These moments are integrals of powers of frequency
multiplied by the corresponding spectral function and integrated over all
frequency. For instance, the spectral moments for the Green's function 
are defined as
$
\mu_m^R = -(1/\pi) \int_{-\infty}^{\infty} d\omega \omega^m \textrm{Im} G(\omega).
$
It can also be shown that 
$
\mu_m^R = \langle \{ [f_i, H]_m, f_i^\dagger \} \rangle,
$
where 
$
\{O_1, O_2 \} = O_1 O_2 + O_2 O_1
$
is the anti-commutator, and $[O_1, O_2]_m$ is the multiple commutation 
operator such that $[O_1, O_2]_1 = [O_1, O_2] = O_1 O_2 - O_2 O_1$;
$[O_1, O_2]_2 = [[O_1, O_2], O_2]$; etc. Evaluating the commutators
is tedious but straightforward, and the results are
\begin{align}
\label{eqn:mu0}
\mu_0^R & = 1, \\
\label{eqn:mu1}
\mu_1^R & = -(\mu_f - U_{bf}\rho_b), \\
\label{eqn:mu2}
\mu_2^R & = \frac{t^{*2}}{2} + (\mu_f - U_{bf}\rho_b)^2 + U_{bf}^2(\langle b_i^\dagger b_i b_i^\dagger b_i \rangle - \rho_b^2), \\
\label{eqn:mu3}
\mu_3^R & = -\frac{3t^{*2}}{2} (\mu_f - U_{bf} \rho_b) - (\mu_f - U_{bf} \rho_b)^3 
- 3\mu_f U_{bf}^2 \nonumber \\ & \times (\langle b_i^\dagger b_i b_i^\dagger b_i \rangle - \rho_b^2) 
+ U_{bf}^3(\langle b_i^\dagger b_i b_i^\dagger b_i b_i^\dagger b_i \rangle - \rho_b^3),
\end{align}
where the operator averages $\langle b_i^\dagger b_i b_i^\dagger b_i \cdots \rangle$ can be easily 
calculated from the knowledge of $w_{n_b}$ as
$
\langle (b_i^\dagger b_i)^k \rangle = \sum_{n_b = 0}^{\infty} w_{n_b} n_b^k,
$
and 
\begin{equation}
\rho_b = \sum_{n_b = 0}^{\infty} w_{n_b} n_b
\end{equation}
is the filling of the bosons. Therefore, the accuracy of the calculations can be further checked
by comparing the real-axis calculation of the moments $\mu_m^R$ by directly 
performing the integrations over the real frequency to the exact 
results given in Eqs.~(\ref{eqn:mu0}) -~(\ref{eqn:mu3}).

Similarly, the spectral moment sum rules for the retarded self-energy are defined as
$
C_m^R = -(1/\pi) \int_{-\infty}^{\infty} d\omega \omega^m \textrm{Im} \Sigma(\omega).
$
One can also calculate these moments by using Dyson's equation, and after some 
significant algebra, the results are
\begin{align}
\label{eqn:C0}
C_0^R &= U_{bf}^2  (\langle b_i^\dagger b_i b_i^\dagger b_i \rangle - \rho_b^2), \\
\label{eqn:C1}
C_1^R &= - U_{bf}^2(\mu_f + 2U_{bf}\rho_b) (\langle b_i^\dagger b_i b_i^\dagger b_i \rangle - \rho_b^2) \nonumber \\
&+ U_{bf}^3 (\langle b_i^\dagger b_i b_i^\dagger b_i b_i^\dagger b_i \rangle - \rho_b^3).
\end{align}
In this way, the accuracy of the calculations can be again benchmarked by comparing 
the real-axis integration of the spectral moments $C_m^R$ with the exact values 
given in Eqs.~(\ref{eqn:C0}) and~(\ref{eqn:C1}). In addition, we would like to 
mention that the large frequency limit of the real-axis self-energy 
\begin{equation}
\Sigma(\omega \to \infty) = U_{bf} \rho_b
\end{equation}
provides another independent check of the numerics.

Typically, there are a number of poles in the self-energy near the correlation 
induced band edges when $U_{bf}$ is large enough, and in order to guarantee the 
accuracy of the sum rules that are calculated on the real-axis by direct quadrature, 
the pole contributions have to be included by hand as described here. 
Using Eqs.~(\ref{eqn:G}) and~(\ref{eqn:G0}) with $i\omega_n \to \omega + i0^+$, 
we find that the locations $\omega_p$ of these poles are determined by the transcendental equation
\begin{equation}
\label{eqn:pole}
\sum_{n_b = 0}^{\infty} \left[w_{n_b} \prod_{m_b \ne n_b}^{\infty} (\omega_p + \mu_f - U_{bf}m_b) \right] = 0.
\end{equation}
In a typical calculation, we restrict the bosonic occupancies to be less 
than some maximal number of multiple boson occupancy. Then the transcendental 
equation becomes a finite equation which is easy to solve directly.
Using such a procedure, the self-energy can then be written as
$
\Sigma(\omega) = \Sigma_{\rm reg}(\omega) + \sum_p R_p/(\omega - \omega_p + i0^+),
$
where $\Sigma_{\rm reg}(\omega)$ is the regular piece, $p$ sums over the 
finite number of poles that were found to solve the truncated equation,
and $R_p$ is  the residue of the pole $\omega_p$ that is calculated next 
(only poles with positive residue are included in the sum as described below).
Since the pole contribution dominates the spectral functions near the poles, 
one can expand Eq.~(\ref{eqn:Gk}) with $i\omega_n \to \omega + i0^+$ in 
powers of $\omega + \mu_f - \Sigma(\omega)$, and obtain
$
G(\omega) \approx g_1/[\omega + \mu_f - \Sigma(\omega)] 
+ g_3/[\omega + \mu_f - \Sigma(\omega)]^3
+ \cdots
$
where $g_1 = \int_{-\infty}^{\infty} d\epsilon \rho(\epsilon) = 1$ and 
$g_3 = \int_{-\infty}^{\infty} d\epsilon \rho(\epsilon) \epsilon^2 = t^{*2}/2$.
Plugging this expansion into Eq.~(\ref{eqn:G0}) to obtain $\Sigma(\omega)$
near the poles, and after some algebra, leads to
\begin{align}
\label{eqn:residue}
& R_p = \frac{\sum_{n_b = 0}^{\infty}(w_{n_b} - \delta_{0n_b}) \prod_{m_b \ne n_b}^{\infty}(\omega_p + \mu_f - U_{bf}m_b)}
{(\omega_p + \mu_f)^{-1} \prod_{q \ne p}(\omega_p - \omega_{q})} \nonumber \\
& - \frac{\sum_{n_b = 0}^{\infty} w_{n_b} \sum_{n_b^\prime \ne n_b}^{\infty} \prod_{m_b \ne \{n_b,n_b^\prime\}}^{\infty}(\omega_p + \mu_f - U_{bf}m_b)}
{g_3^{-1}\prod_{q \ne p}(\omega_p - \omega_{q})},
\end{align}
where $\delta_{ij}$ is the Kronecker delta function. Only poles where the residue 
$R_p$ is positive correspond to real physical poles. As a result, the pole 
contribution to the self-energy moment $C_m^R$ is $\sum_{p'} R_{p'} \omega_{p'}^m$, 
where $p'$ sums over the poles with positive residues. 

We recall that the spectral moment sum rules given in Eqs.~(\ref{eqn:mu0}) -~(\ref{eqn:mu3}) 
and Eqs.~(\ref{eqn:C0}) and~(\ref{eqn:C1}) have exactly the same 
form in the case of the Fermi-Fermi FK model~\cite{sum1,sum2,sum3}. 
However, in that case 
$
\langle (b_i^\dagger b_i)^k \rangle = \langle b_i^\dagger b_i \rangle = w_1
$
for any $k$ due to the fermionic anticommutators. In addition, 
since $w_{n_b \ge 2} = 0$, the pole equation given in Eq.~(\ref{eqn:pole}) 
reduces considerably leading to a single pole at 
$\omega_1 = -\mu_f + U_{bf} (1-w_1)$, for which the residue 
equation given in Eq.~(\ref{eqn:residue}) reduces to the well-known 
result $R_1 = w_1(1-w_1) U_{bf}^2 - g_3$~\cite{prl_demchenko}.

\subsection{Asymptotic Behavior for Large Frequencies}
\label{sec:methods}

There is another important application of the spectral moment sum rules given above. 
They can be used to evaluate the high-frequency asymptotic behavior of the 
Green's function, self-energy and dynamical mean-field exactly. Therefore, this 
knowledge can be used to reduce the number of Matsubara frequencies 
that is needed to solve the imaginary-axis equations~\cite{sum3}. 

When the Matsubara frequency is high enough, it can be shown that 
the Green's function can be written as
$
G(i\omega_n) = \sum_{m = 0}^{\infty} \mu_m^R/(i\omega_n)^{m+1},
$
and similarly the self-energy can be written as
$
\Sigma(i\omega_n) = \Sigma(\infty) + \sum_{m = 0}^{\infty} C_m^R/(i\omega_n)^{m+1},
$
where $\Sigma(\infty)$ is a real constant, and it is the large-frequency limit
of the self-energy. These expansions follow from the definition of the 
spectral moment sum rules and the spectral formula for the retarded Green's function and self-energy.
Plugging these expansions into Eq.~(\ref{eqn:G0}), and using the definition
of $G_0^{-1}(i\omega_n)$ given below Eq.~(\ref{eqn:G}), we obtain
the asymptotic expansion of the dynamical mean-field as
$
\lambda(i\omega_n) = t^{*2}/[2(i\omega_n)] - t^{*2}(\mu_f - U_{bf} \rho_b)/[2(i\omega_n)^2] + \cdots.
$
This asymptotic expansion along with the expansion above allow us to 
treat the high-frequency tails of some quantities as described next.

Let's take an energy cutoff $\epsilon_c$ which is much larger than the bandwidth of
the interacting DOS, and use the asymptotic expansion to sum over
the Matsubara frequencies that are higher than $\epsilon_c$. This also defines a 
cutoff $n_c$ for the Matsubara frequencies, given by the closest one to $\epsilon_c$
but lying below it, \textit{i.e.} $\omega_{n_c} = (2n_c+1)\pi/\beta \lesssim \epsilon_c$.  
Plugging the asymptotic expansions into Eq.~(\ref{eqn:Znb}), we obtain
\begin{align}
&Z_{n_b} = 2 \exp \left \{\beta\left[\frac{\mu_f - U_{bf} n_b}{2} + \mu_b n_b - \frac{U_{bb}}{2}n_b(n_b-1) \right]\right \} \nonumber \\
& \times \prod_{n = 0}^{n_c} \left| \frac{G_0^{-1}(i\omega_n) - U_{bf} n_b}{i\omega_n} \right|^2
\prod_{n = n_c+1}^{\infty} \left| 1 + \frac{\mu_f - U_{bf} n_b}{i\omega_n} \right. \nonumber \\
&\left. - \frac{t^{*2}}{2(i\omega_n)^2} 
+ \frac{t^{*2}(\mu_f - U_{bf}\rho_b)}{2(i\omega_n)^3} \right|^2.
\end{align}
Here, since the complex conjugate of $i\omega_n$ is equal to $i\omega_{-n-1}$, we take the 
negative Matsubara frequencies into account by writing the absolute value squares 
in the products. To approximate the infinite products that have an infinite number of 
terms, we first rewrite the infinite products as the exponential of the sum of 
the logarithm of the individual terms. Then, for temperatures much lower than 
the bandwidth of the interacting DOS, we replace the sum by
an integral, and then convert the integral over frequency to an integral 
over $z = 1/\omega$, leading to
\begin{align}
\prod_{n = n_c+1}^{\infty} &\left| 1 + \frac{a}{i\omega_n} + \frac{b}{(i\omega_n)^2} + \frac{c}{(i\omega_n)^3} \right|^2
\approx \exp\Big\lbrace  \frac{\beta}{2\pi} \nonumber \\ 
& \times \int_0^{\frac{1}{\epsilon_c}} \frac{dz}{z^2} \ln\left[ (1 + bz^2)^2 + z^2(a - cz^2)^2 \right] \Big\rbrace.
\end{align}
Here, $a = \mu_f - U_{bf} n_b$, $b = -t^{*2}/2$, and $c = t^{*2}(\mu_f - U_{bf}\rho_b)/2$.
As a result, the use of asymptotic expressions for the Green's function, self-energy, 
and the dynamical mean-field allow us to reduce the computational effort 
considerably for the imaginary-axis calculation by keeping $\epsilon_c$ small. 
Because the DOS extends farther out in energy as $T$ rises, one needs to 
increase $\epsilon_c$ in order to achieve the same level of accuracy for high $T$.

The asymptotic expansions can also be used to treat the tails in the summation
for the fermion filling, \textit{i.e.} $\rho_f = (1/\beta) \sum_{n=-\infty}^{\infty} G(i\omega_n)$,
which leads to
\begin{align}
\rho_f = \frac{\mu_0^R}{2} &- \frac{\beta \mu_1^R}{4} + \frac{\beta^3 \mu_3^R}{48} 
+ \frac{1}{\beta} \sum_{n = -n_c}^{n_c-1} \left[ G(i\omega_n) \right. \nonumber \\ 
&\left. - \frac{\mu_0^R}{i\omega_n} - \frac{\mu_1^R}{(i\omega_n)^2} - \frac{\mu_2^R}{(i\omega_n)^3} - \frac{\mu_3^R}{(i\omega_n)^4} \right].
\end{align}
Here, we use the standard representation of the Fermi-Dirac distribution function
$
F(x) = (1/\beta) \sum_{n = -\infty}^{\infty} 1/(i\omega_n - x)
$ 
and its derivatives with respect to $x$ when $x \to 0$, to evaluate the coefficients 
of the $(1/\beta) \sum_{n = -\infty}^{\infty} 1/(i\omega_n)^k$ type. 
Since the contribution from the summation (which needs to be evaluated 
numerically) vanishes rapidly for large $\omega_n$, the filling can be 
calculated quite accurately.

\begin{figure*} [htb]
\centerline{\scalebox{0.45}{\includegraphics{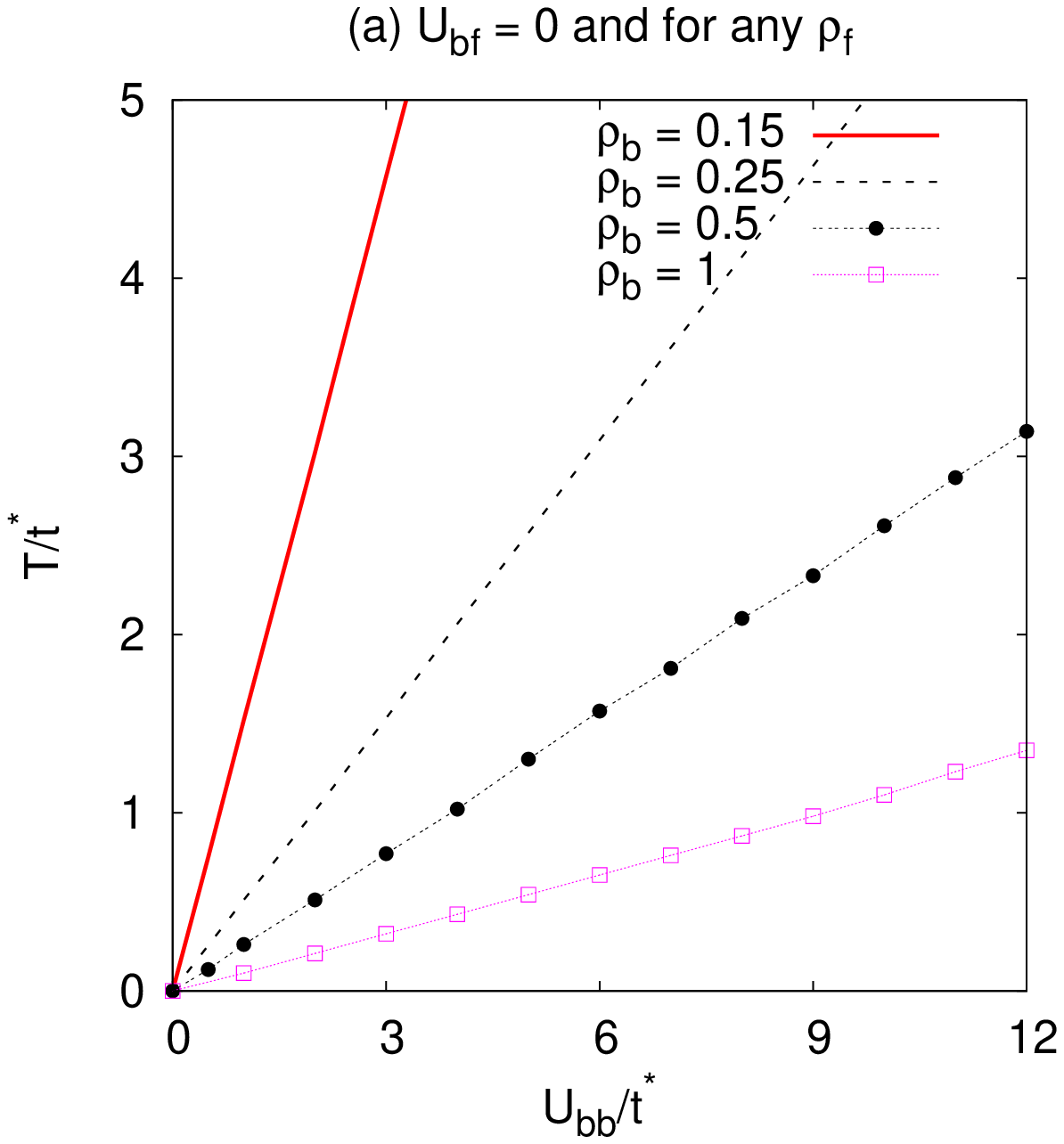}, \includegraphics{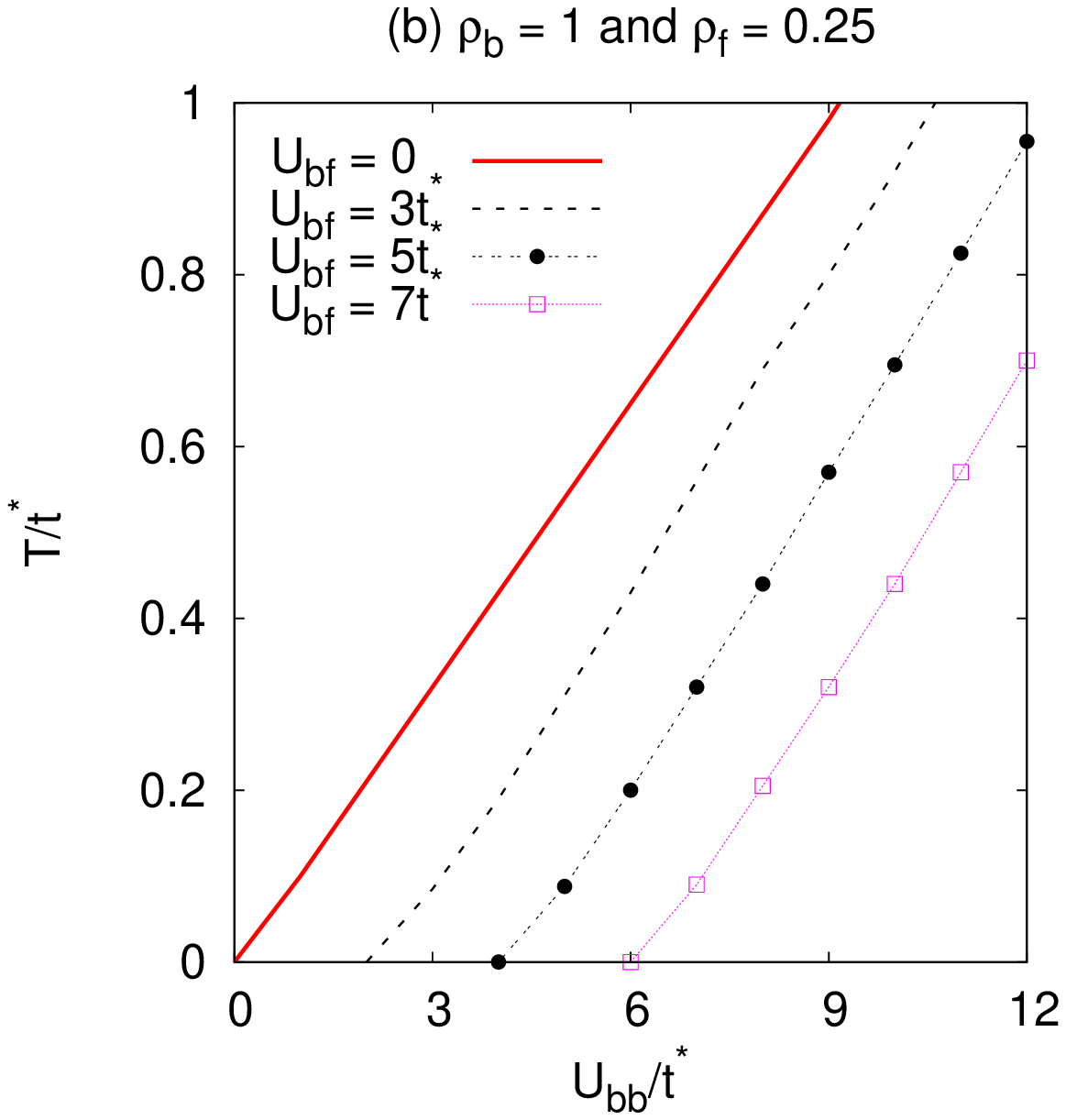}, \includegraphics{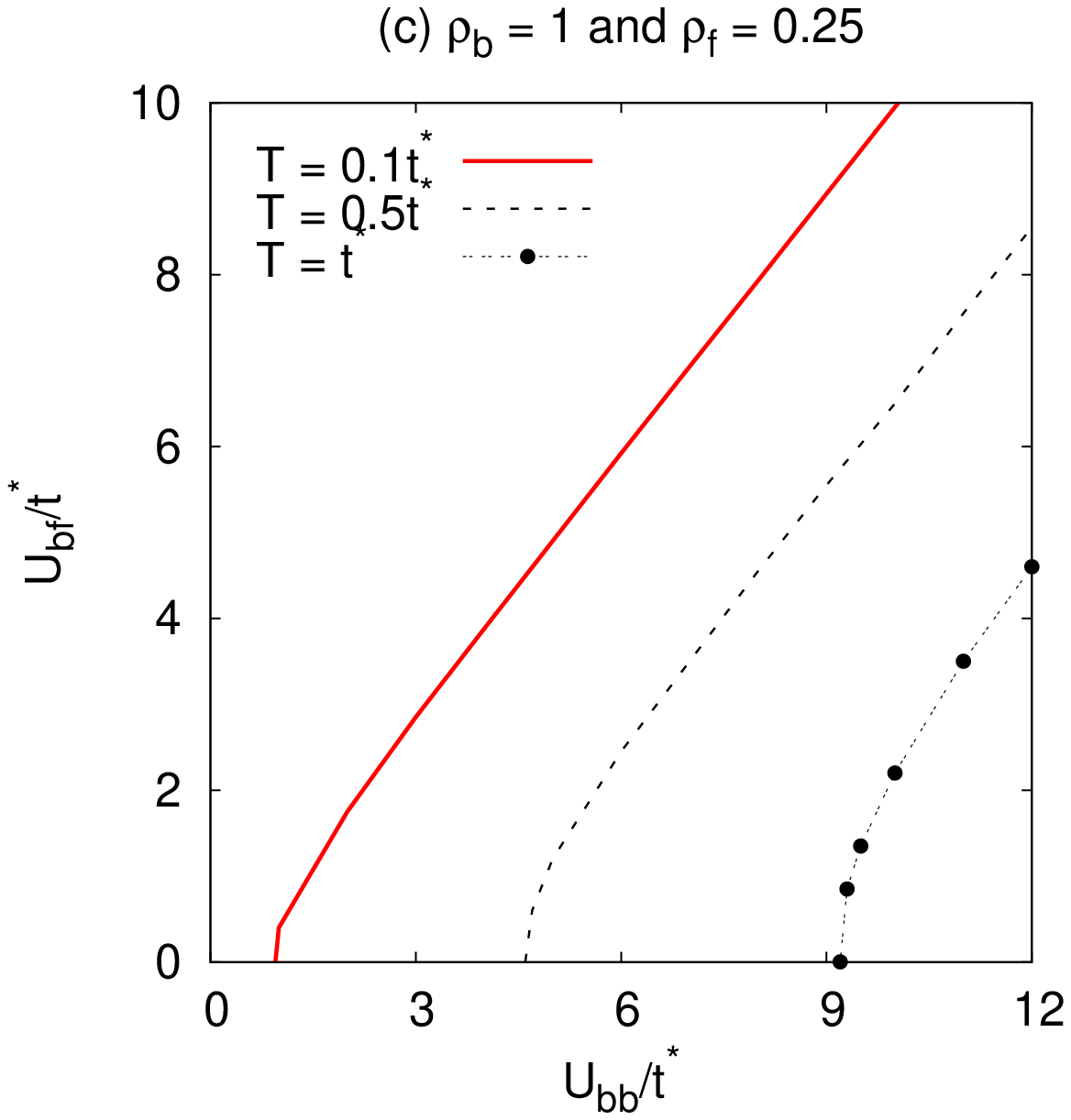}}}
\caption{\label{fig:mapping}
Mapping diagrams show when the Fermi-Bose FK model can be mapped onto
the spinless Fermi-Fermi FK model as a function of boson-boson repulsion $U_{bb}$. 
In all figures, mapping boundaries separate the regions where total probability 
of finding a boson in two of the $n_b$ states is higher than $99\%$
(which occurs below each line).
}
\end{figure*}
\begin{figure*} [htb]
\centerline{\scalebox{0.45}{\includegraphics{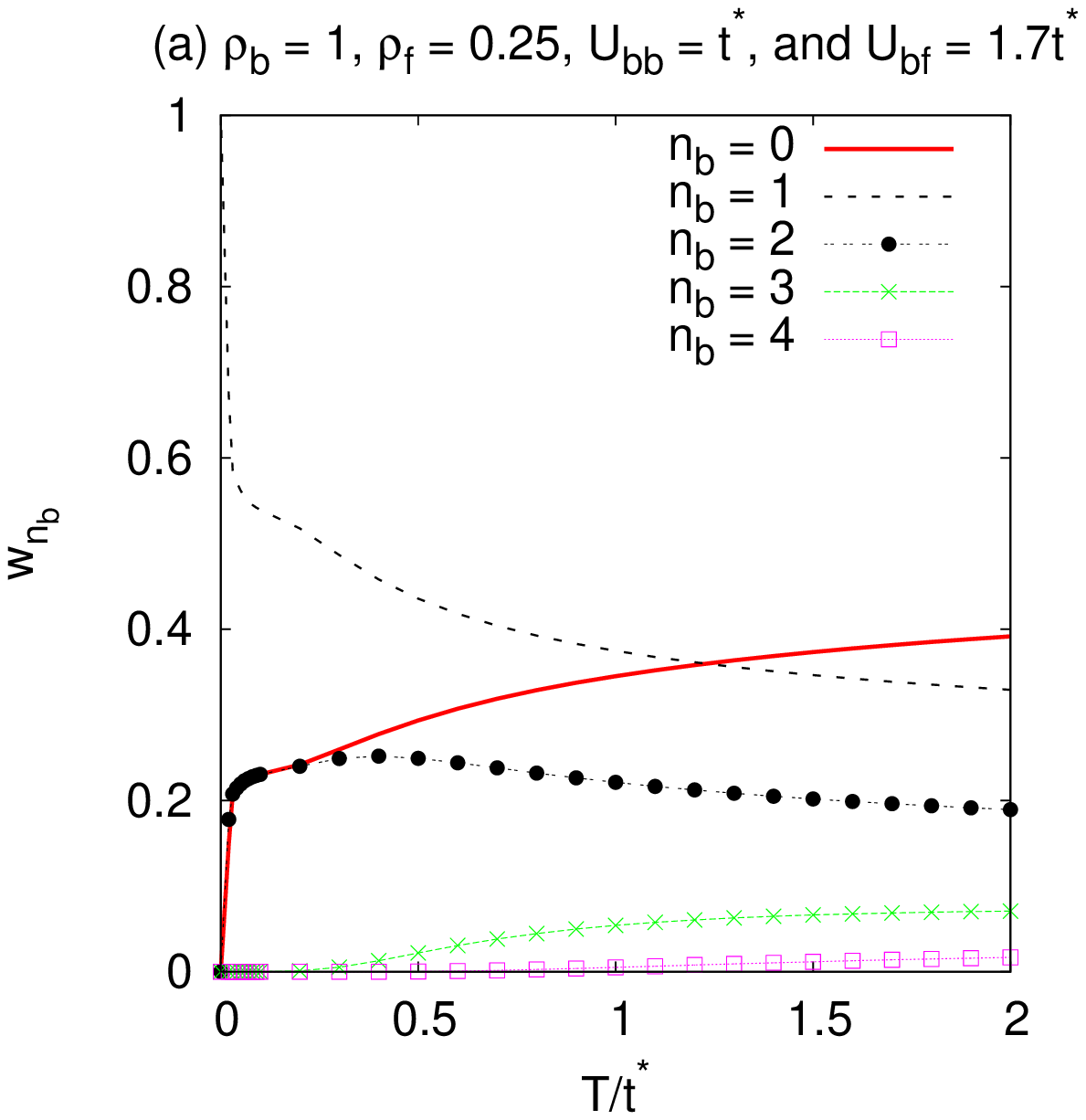}, \includegraphics{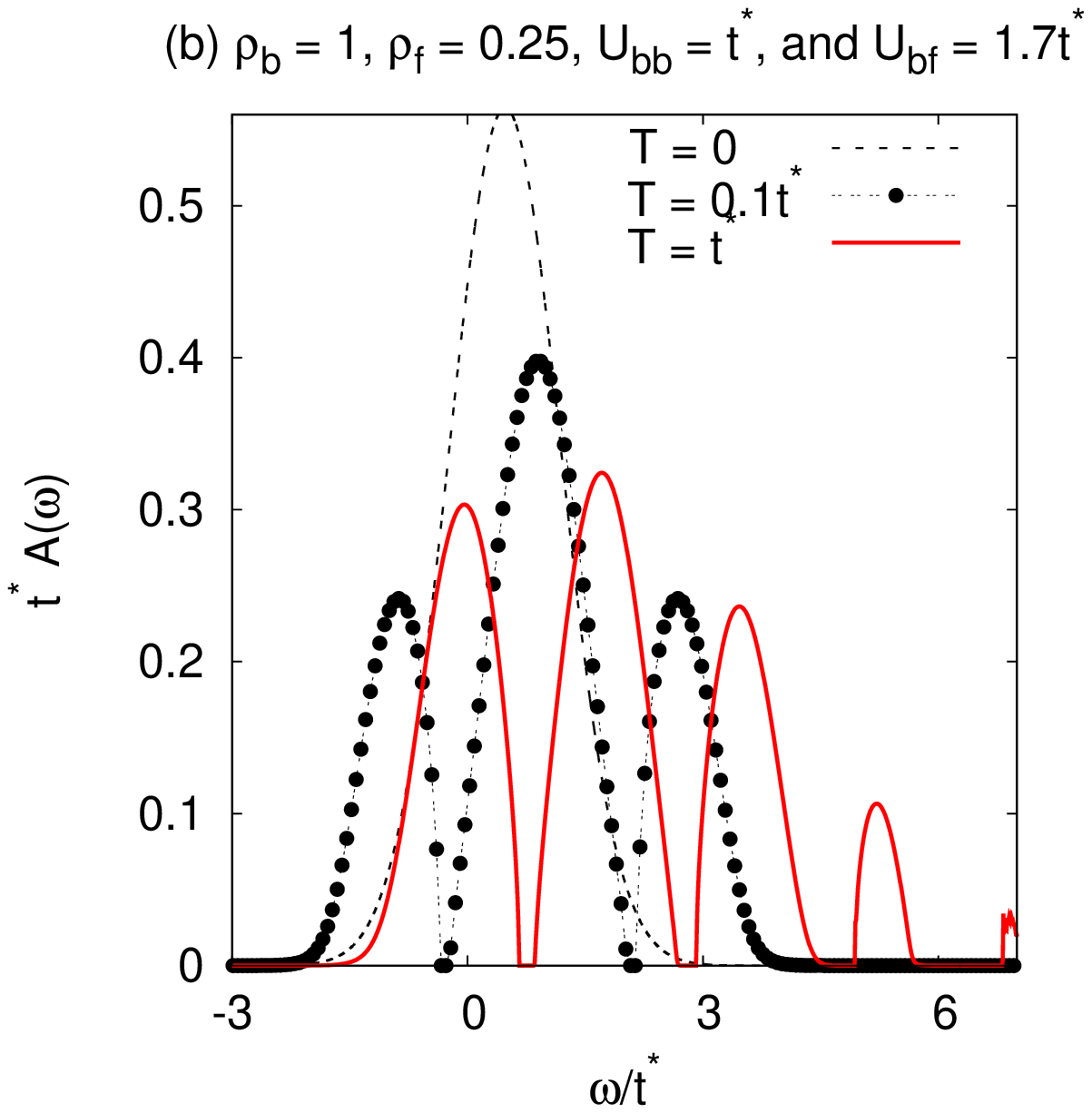}, \includegraphics{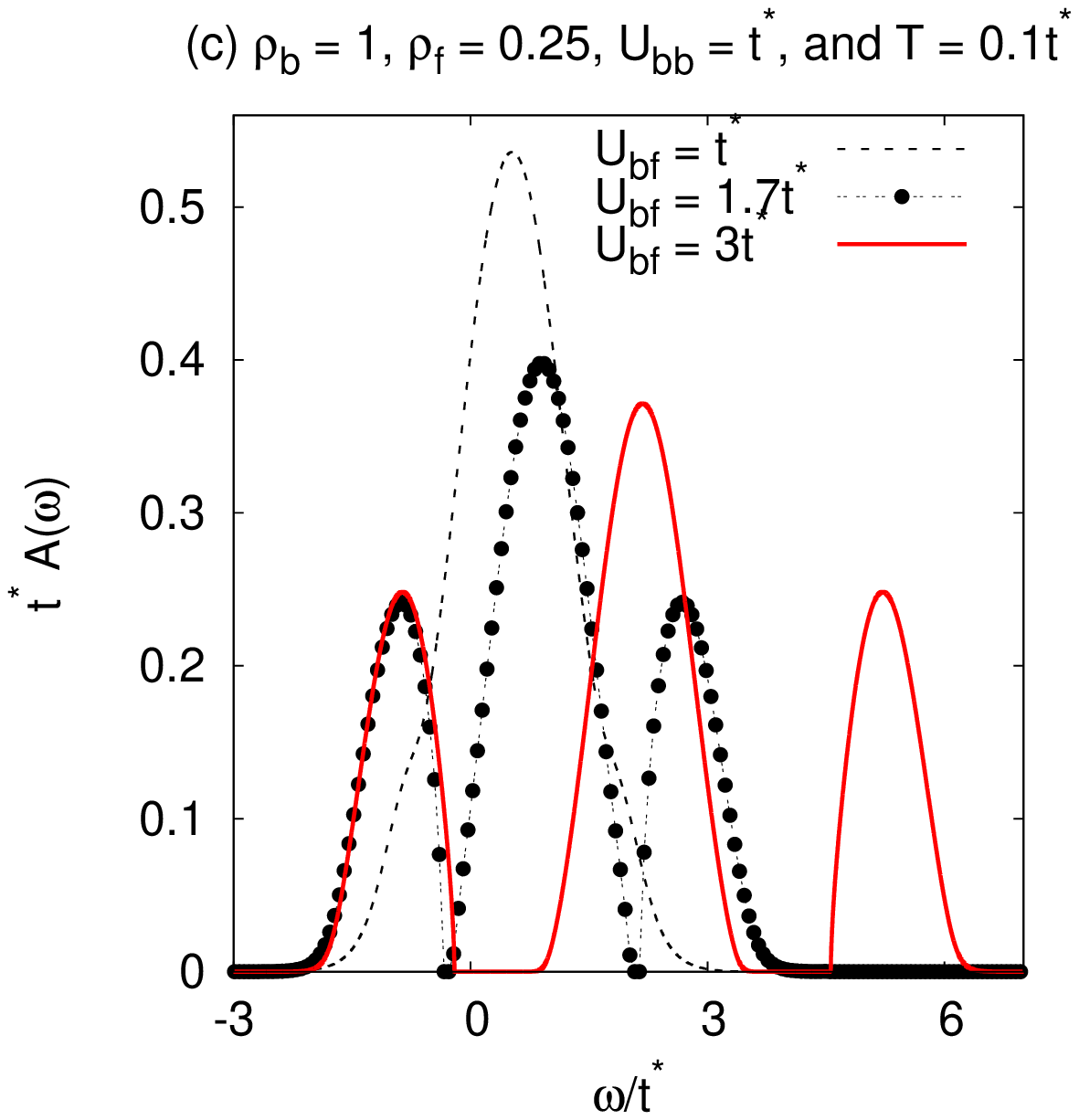}}}
\caption{\label{fig:dos}
(a) The probability $w_{n_b}$ of finding a boson in state $n_b$ is shown as a function 
of temperature $T$ for the first five boson occupancies.
(b, c) The single-particle many-body density of states for the fermions $A(\omega)$ is shown 
as a function of frequency $\omega$. 
}
\end{figure*}
\section{Numerical Results}
\label{sec:numerics}

Having described the DMFT formalism for the Fermi-Bose FK model, next 
we present some of our numerical results that are obtained by solving
Eqs.~(\ref{eqn:G}),~(\ref{eqn:G0}) and~(\ref{eqn:Gk}) self-consistently 
for given fermion and boson fillings, $\rho_f$ and $\rho_b$, respectively. 
For this purpose, we choose a large energy cutoff $\epsilon_c = 200t^*$, 
and treat the tails of the high-frequency summations using the formalism 
described above. Since the system is unstable for $U_{bb} < 0$, 
and the $U_{bf} < 0$ case can be mapped onto the $U_{bf} > 0$ case 
as discussed in Sec.~\ref{sec:symmetry}, we restrict our analysis to 
$\{U_{bb}, U_{bf} \} > 0$ values. We remark here that numerical solutions 
with $w_{n_b \to \infty} \ne 0$ are unphysical, indicating an instability 
in the system corresponding to a bosonic collapse when $U_{bb} < 0$.

\subsection{Mapping onto the Spinless Fermi-Fermi FK Model}
\label{sec:mapping}

We argued in Sec.~\ref{sec:symmetry} that the Fermi-Bose FK model 
can be exactly mapped onto the well-studied spinless Fermi-Fermi FK model 
for all parameter space at $T = 0$ as long as the mixture is thermodynamically stable. 
Here, we numerically show how the system evolves away from 
the Fermi-Fermi behavior at higher temperatures.

In Fig.~\ref{fig:mapping}, we calculate the region where the total probability 
of finding a boson in just two of the $n_b$ states is higher than $99\%$. 
Therefore, below each line in all figures, the system can be effectively 
mapped onto the spinless Fermi-Fermi FK model. In Fig.~\ref{fig:mapping}(a), 
we show temperature $T$ vs. boson-boson repulsion $U_{bb}$ mapping diagram 
for different boson fillings $\rho_b$ when bosons and fermions 
are uncoupled, \textit{i.e.} $U_{bf} = 0$.
As $\rho_b$ increases, it is seen that the mapping is possible only for
lower $T$ values at a given $U_{bb}$, and is possible only for lower $U_{bb}$ 
values at a given $T$. This is because it is energetically more favorable to 
have an occupation of multiple $n_b$ states as a function of increasing $T$ 
and/or decreasing $U_{bb}$ due to the Bose distribution function. 

When $U_{bf} \ne 0$, we show a $T$ vs. $U_{bb}$ mapping diagram for different 
$U_{bf}$ in Fig.~\ref{fig:mapping}(b), and a $U_{bf}$ vs. $U_{bb}$ 
mapping diagram for different $T$ in Fig.~\ref{fig:mapping}(c).
As $U_{bf}$ increases, it is seen that the mapping is possible for smaller 
and smaller parameter space compared to the $U_{bf} = 0$ limit.
This is because the coupling between bosons and fermions induces an 
attractive interaction between bosons such that the effective boson-boson 
repulsion $U_{bb}^{\rm eff}$ decreases by some amount that is proportional 
to $U_{bf}^2$ to the lowest order in $U_{bf}$. 
Therefore, increasing $U_{bf}$ leads to an occupation of additional $n_b$ states
as discussed next.

\subsection{Density of States (DOS) for the Fermions}
\label{sec:dos}

In this subsection, we present our numerical results for the probability $w_{n_b}$ 
of finding a boson in state $n_b$, and the single-particle many-body density of states 
(DOS) for the fermions, to illustrate typical properties of the Fermi-Bose FK model. 
The probabilities are shown in Fig.~\ref{fig:dos}(a) as a function of $T$ for the
first five boson occupancies when $\rho_b = 1$, $\rho_f  = 0.25$, $U_{bb} = t^*$, 
and $U_{bf} = 1.7t^*$. It is seen that $w_{n_b} = \delta_{1n_b}$ at $T = 0$
which is due to $\rho_b = \sum_{n_b = 0}^{\infty} w_{n_b} n_b$ with 
$\sum_{n_b = 0}^{\infty} w_{n_b} = 1$. 
However, the occupation of the $n_b = 1$ state decreases at finite $T$, 
while that of higher and higher $n_b$ states become finite with increasing $T$. 

The occupancy of multiple $n_b$ states has a strong effect on the many-body DOS 
for the fermions, \textit{i.e.} 
$
A(\omega) = -(1/\pi) \mathrm{Im} G(i\omega_n \to \omega + i0^+),
$ 
and is given by
\begin{equation}
A(\omega) = -\frac{1}{\pi} \mathrm{Im} \int_{-\infty}^{\infty} 
\frac{d\epsilon \rho(\epsilon)}{\omega + \mu_f - \Sigma(\omega) -\epsilon + i0^+},
\end{equation}
where $\rho(\epsilon)$ is the noninteracting DOS for the infinite-dimensional hypercubic
lattice defined below Eq.~(\ref{eqn:Gk}),
and the infinitesimal is needed only when $\mathrm{Im} \Sigma(\omega) = 0$. 
In Fig.~\ref{fig:dos}(b), there is a single Gaussian peak at $T = 0$
corresponding to $n_b = 1$ (which actually becomes a noninteracting system), 
but there are three peaks at $T = 0.1t^*$
corresponding to $n_b = 0, 1$ and 2. In addition, two more peaks occur 
at $T = t^*$ corresponding to $n_b = 3$ and 4. This is again because 
it is energetically more favorable to have an occupation of multiple $n_b$ 
states as a function of increasing $T$ due to the bosonic character of the heavy atoms. 
We mention that while there is not any pole in the self-energy when
$T = 0$, there are two physical poles at $\omega_1 \approx -0.348t^*$
and $\omega_2 \approx 2.149t^*$ with residues $R_1 \approx R_2 \approx 0.166t^*$
when $T = 0.1t^*$, and at $\omega_1 \approx 0.725t^*$ and $\omega_2 \approx 2.871t^*$
with residues $R_1 \approx 0.419t^*$ and $R_2 \approx 0.402t^*$ when $T = t^*$.
Notice that poles occur near the correlation induced band edges.

We investigate the effects of $U_{bf}$ on $A(\omega)$ in Fig.~\ref{fig:dos}(c),
where it is seen that increasing (decreasing) $U_{bf}$ increases (decreases) 
the number of peaks. This is again because the coupling between the
bosons and fermions induces an attractive interaction between bosons
such that the effective boson-boson repulsion decreases, which
leads to an occupation of additional $n_b$ states. For instance, as shown in the figure, 
there is a single peak when $U_{bf} = t^*$, but there are three of them 
when $U_{bf} = 1.7t^*$. In addition, the band gap between different 
$n_b$ states increases with increasing $U_{bf}$. 
We again mention that while there is not any pole in the self-energy when
$U_{bf} = t^*$, there are two physical poles at $\omega_1 \approx -0.348t^*$
and $\omega_2 \approx 2.149t^*$ with residues $R_1 \approx R_2 \approx 0.166t^*$
when $U_{bf} = 1.7t^*$, and at $\omega_1 \approx 0.0605t^*$ and $\omega_2 \approx 4.295t^*$
with residues $R_1 \approx R_2 \approx 1.759t^*$ when $U_{bf} = 3t^*$.
Notice again that poles occur near the correlation induced band edges.

Although the DOS for the fermions shows rich structures, this is not yet a 
measurable quantity in ultracold atomic systems (it might be 
feasible with an appropriately designed rf-frequency experiment that 
acts like a photoemission experiment of the fermions~\cite{jin}, 
but there one would be observing the DOS multiplied by the Fermi-Dirac 
distribution function, the so-called lesser spectral function).
For this reason, we next discuss the momentum distribution of the
fermions and bosons which could be easily measured in a time-of-flight measurement.

\begin{figure} [htb]
\centerline{\scalebox{0.45}{\includegraphics{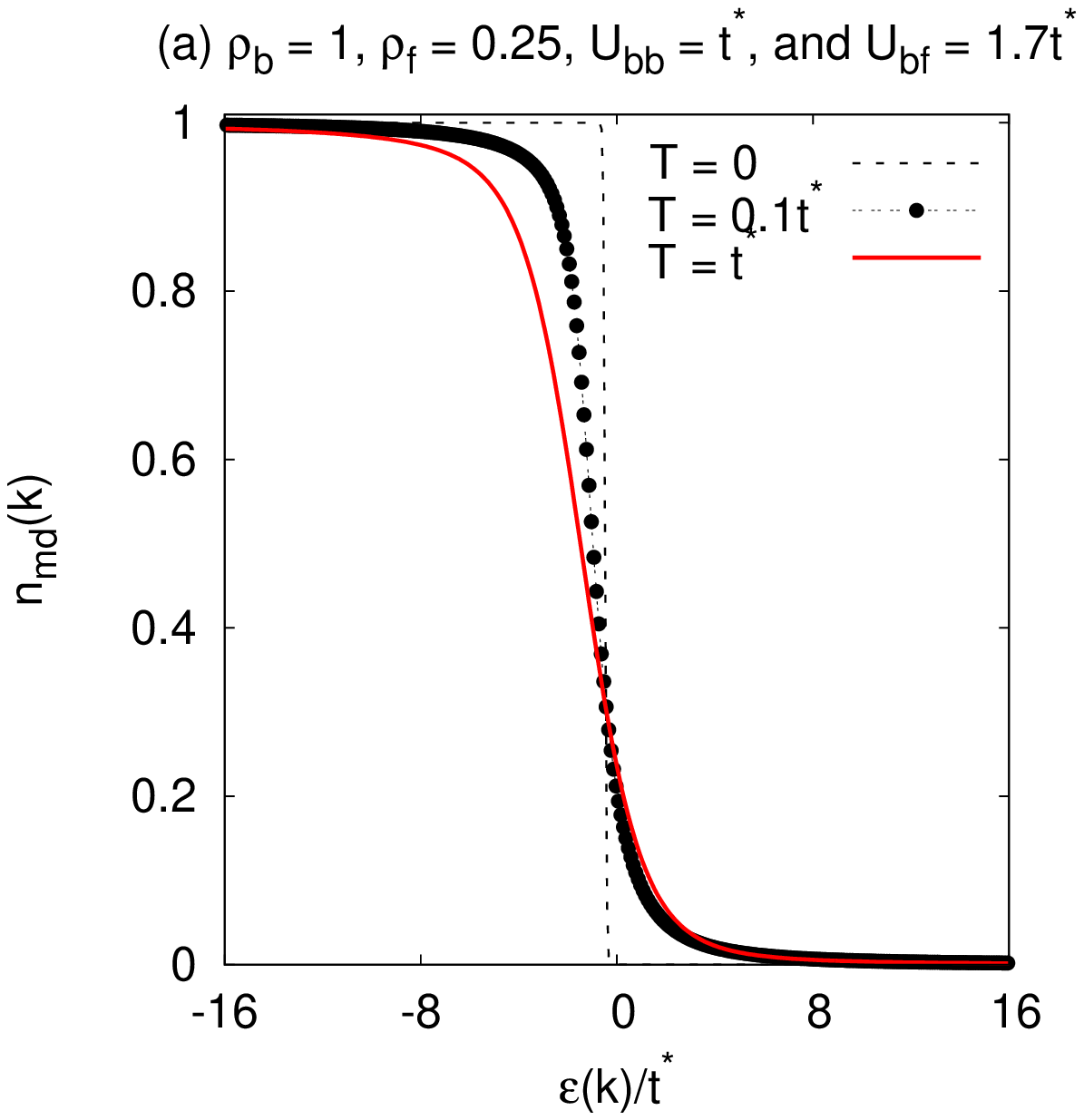}}}
\centerline{\scalebox{0.45}{\includegraphics{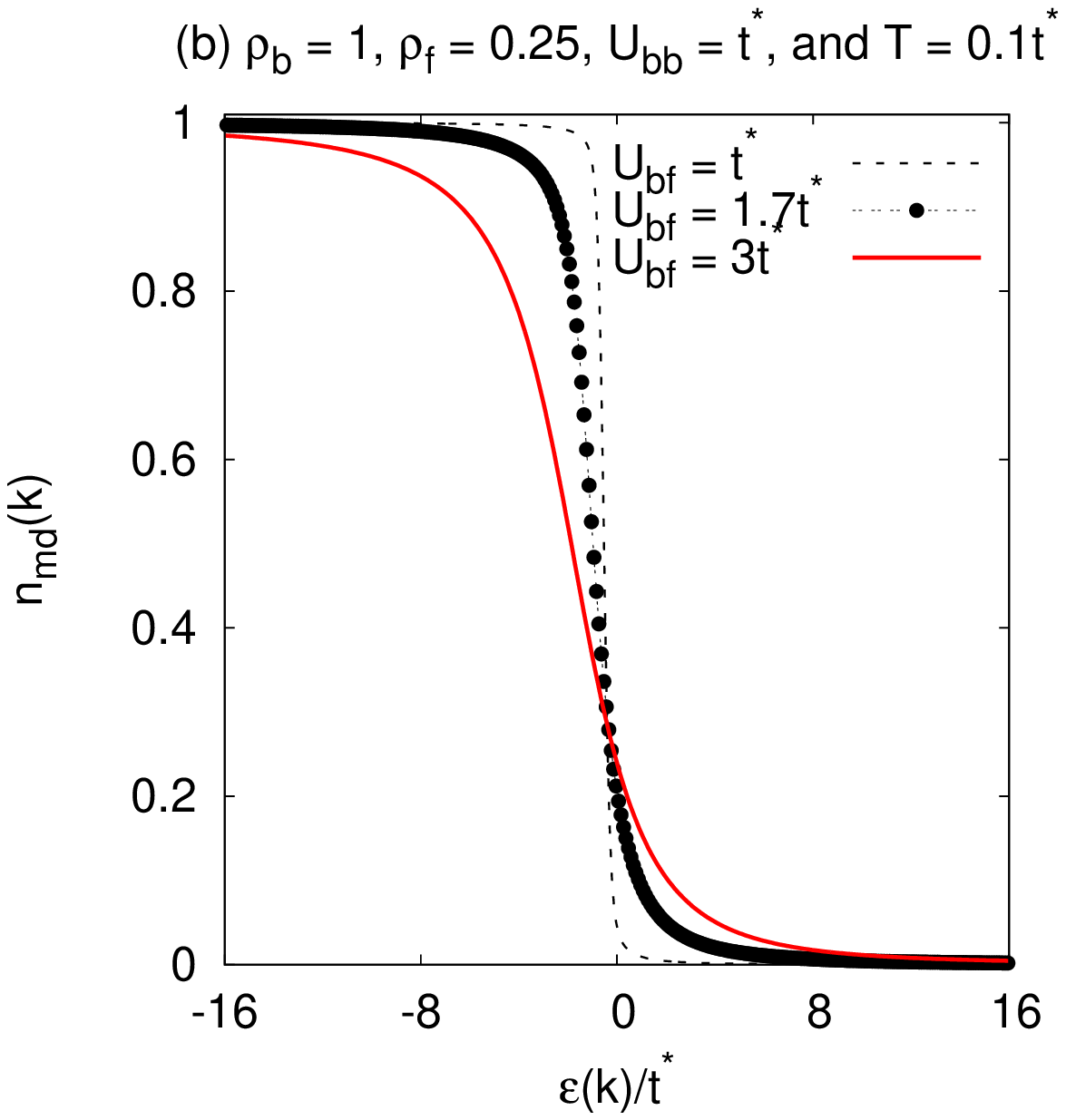}}}
\caption{\label{fig:md}
The momentum distribution of the fermions $n_{md}(\mathbf{k})$ is shown 
as a function of the dispersion $\epsilon(\mathbf{k})$. 
The parameters used in these figures are the same as the ones 
used in Fig.~\ref{fig:dos}.
}
\end{figure}
\subsection{Momentum Distribution}
\label{sec:md}

The occupancy of multiple $n_b$ states has also significant effect 
on the momentum distribution of the fermions, \textit{i.e.}
$n_{md}(\mathbf{k}) = (1/\beta) \sum_{n = -\infty}^{\infty} G(\mathbf{k}, i\omega_n)$,
which becomes
\begin{equation}
n_{md}(\mathbf{k}) = \frac{1}{\beta} \sum_{n = -\infty}^{\infty} \frac{1}{i\omega_n + \mu_f - \Sigma(i\omega_n) - \epsilon(\mathbf{k})},
\label{eqn:md}
\end{equation}
where $G(\mathbf{k}, i\omega_n)$ is the momentum-resolved Green's function
and $\epsilon(\mathbf{k}) = -2t_f\sum_{i = 1}^{d} \cos(k_i a)$ is the energy dispersion
for the fermions defined below Eq.~(\ref{eqn:Gk}). 
The momentum distribution of the bosons is simply a constant given by the filling 
of the bosons since they are localized with zero tunneling. 
To regularize the frequency summation in our numerical calculations, we subtract
$
(1/\beta) \sum_{n = -\infty}^{\infty} 1/[i\omega_n + \mu_f - \Sigma(\infty) - \epsilon(\mathbf{k})],
$
and add $F[-\mu_f + \Sigma(\infty) + \epsilon(\mathbf{k})]$ 
in Eq.~(\ref{eqn:md}), where $F(x)$ is the Fermi-Dirac distribution function. 

Our numerical results for $n_{md}(\mathbf{k})$ vs. the fermion dispersion 
$\epsilon(\mathbf{k})$ are shown in Fig.~\ref{fig:md}. As shown in Fig.~\ref{fig:md}(a), 
$n_{md}(\mathbf{k})$ is a step function at $T = 0$, and it broadens as a 
function of increasing $T$, which is expected due to Fermi-Dirac statistics.
For a fixed $T$, the effects of $U_{bf}$ on $n_{md}(\mathbf{k})$ 
are shown in Fig.~\ref{fig:md}(b), where it is seen that increasing 
$U_{bf}$ also broadens $n_{md}(\mathbf{k})$ just like the temperature, 
which is the expected many-body effect due to a finite lifetime of 
the fermionic excitations. In the FK model, the fermions are usually not 
a Fermi liquid (or more correctly a Fermi gas in our context) at $T = 0$, 
but for the case of integer boson fillings $\rho_b = 0, 1, 2, \cdots, \infty$ 
(presented here) they are, because the system evolves to an effective 
noninteracting system as $T \rightarrow 0$.

\subsection{Average Kinetic Energy}
\label{sec:ake}

Another important quantity that can be measured in ultracold atomic 
systems is the average kinetic energy of the particles, which is given by
$
\langle \epsilon(\mathbf{k}) \rangle = \sum_{\mathbf{k}} \epsilon(\mathbf{k}) n_{md}(\mathbf{k})
$
for the fermions, and vanishes for the bosons since they are localized 
with zero tunneling. Since the momentum distribution is measured directly 
in experiment, one can process the data to determine the average kinetic energy 
in the lattice under the assumption that the momentum distribution has not 
changed significantly during the time-of-flight experiment. For our calculations, 
we convert the summation over momentum to an integral over energy. 
The resulting expression is
\begin{equation}
\langle \epsilon(\mathbf{k}) \rangle = \int_{-\infty}^{\infty} d\epsilon \rho(\epsilon) \epsilon n_{md}(\epsilon),
\end{equation}
where $\rho(\epsilon)$ is the noninteracting DOS for the infinite-dimensional hypercubic
lattice defined below Eq.~(\ref{eqn:Gk}).

\begin{figure} [htb]
\centerline{\scalebox{0.45}{\includegraphics{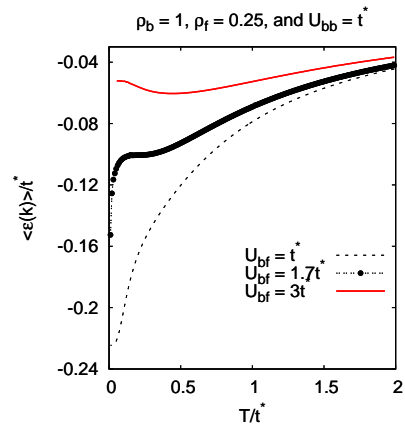}}}
\caption{\label{fig:ake}
The average kinetic energy of the fermions $\langle \epsilon(\mathbf{k}) \rangle$ 
is shown as a function of temperature $T$. 
}
\end{figure}

Our numerical results for $\langle \epsilon(\mathbf{k}) \rangle$ vs. $T$ 
are shown in Fig.~\ref{fig:ake}. When $U_{bf}$ is small, \textit{e.g.} $U_{bf} = t^*$, 
it is seen that $\langle \epsilon(\mathbf{k})\rangle$ increases monotonically 
as a function of $T$. However, when $U_{bf}$ becomes large enough, 
see \textit{e.g.} $U_{bf} = 1.7t^*$, it has a local minimum at 
finite $T$ after an initial increase for lower temperatures. 
This minimum moves towards higher $T$ as $U_{bf}$ becomes larger, 
compare \textit{e.g.} $U_{bf} = 1.7t^*$ case with that of $U_{bf} = 3t^*$.
The nonmonotonic behavior occurs only for large enough $U_{bf}$ values and
is a consequence of the correlation induced band gaps that are present 
in the many-body DOS, as discussed in Sec.~\ref{sec:dos}. This may be one
of the most direct ways, with current available technology, to infer the
changes in the DOS as a function of $T$ in experiment.

\section{Conclusions}
\label{sec:conclusions}

In this work, we analyzed Fermi-Bose mixtures consisting of 
light fermions and heavy bosons that are loaded into 
optical lattices. To describe such mixtures, we considered 
the Fermi-Bose version of the FK model, the Fermi-Fermi version 
of which has been widely discussed in the condensed matter 
literature. In our model, we assumed that the bosons are 
localized such that their tunneling to other sites vanishes 
($t_b = 0$) but that the system can statistically sample all low 
energy configurations of the heavy atoms. This perspective makes 
sense in ultracold atomic experiments if the difference in the tunneling 
amplitudes between the light fermions and heavy bosons is large 
enough so that the quantum nature of the bosons can be neglected, 
but they can reorganize their positions to allow the system 
to sample different configurations. An alternative perspective 
is that once the optical lattice has been introduced, the heavy 
bosons become frozen into a specific configuration, that is 
randomly chosen from the configurations that are energetically 
favorable in the statistical mechanical ensemble; as one repeats 
many experiments and averages over the different configurations, 
one would then reproduce the results of the FK model described here~\cite{patterns}.

First, we discussed the symmetries of the Hamiltonian, and showed that 
the Fermi-Bose FK model can be mapped exactly onto the spinless 
Fermi-Fermi FK model at zero temperature for all parameter space
as long as the mixture is thermodynamically stable. 
Since this mapping is only approximate at low temperatures and 
fails at high temperatures, we developed DMFT to investigate the 
effects of temperature (recall that the DMFT becomes exact in infinite dimensions). 
We also calculated spectral moment sum rules for the retarded Green's function 
and retarded self-energy, and used them to check the accuracy of our numerical 
calculations, as well as to reduce the computational cost. 
When bosons and fermions are uncoupled ($U_{bf} = 0$), we showed that, 
as the boson filling increases, the mapping is possible only for
lower $T$ values at a given $U_{bb}$, and is possible only for lower $U_{bb}$ 
values at a given $T$. This is because it is energetically more favorable to 
have occupation of multiple $n_b$ states as a function of increasing $T$ 
and/or decreasing $U_{bb}$ due to the Bose statistics. 
As $U_{bf}$ increases, we found that the mapping is possible for smaller 
and smaller parameter space compared to the $U_{bf} = 0$ limit.
This is because the coupling between bosons and fermions induces an 
attractive interaction between bosons such that the effective boson-boson 
repulsion $U_{bb}^{\rm eff}$ decreases by some amount that is proportional to 
$U_{bf}^2$ to the lowest order in $U_{bf}$. Therefore, increasing $U_{bf}$ 
leads to an occupation of additional $n_b$ states.

We also presented typical numerical results for the Fermi-Bose FK model 
including the occupancy of bosonic states, single-particle many-body DOS
for the fermions, experimentally relevant momentum distribution, and
the average kinetic energy. 
We found that the occupancy of multiple bosonic states has a strong 
effect on the DOS for the fermions, leading to strong 
modulations as a function of frequency. The number of peaks 
corresponds to the number of bosonic states that are occupied, and it 
increases as a function of increasing $T$ and/or $U_{bf}$. 
In addition, we showed that increasing $U_{bf}$ at a fixed $T$ 
broadens the momentum distribution of the fermions, just like the 
effects of temperature by itself. We also showed how the average kinetic 
energy evolves with $T$ and how one can infer changes to the DOS 
via structure in the average kinetic energy.

We hope that some of these results could be experimentally realized in 
ultracold atomic systems. We think, for instance, 
K-Rb, Li-K, or Li-Cs mixtures are
a good initial candidates for simulating the Bose-Fermi FK model. In addition, 
one could also create species-dependent optical lattices for different 
isotopes of the same atom such that the bosonic isotope is localized
but not the fermionic one. 

Finally, if one recalls the local-density approximation as a first 
approximation to the effects of the trap in a real experimental system, 
we expect that the methods described here could be quickly used 
to generate approximate results for density distributions across 
the trap and for a variation of the density of states or of the 
momentum distribution. Such results are beyond the scope of this work, 
but could be investigated if one is interested in directly modeling 
a specific experiment that is described by the Fermi-Bose FK model.

\acknowledgments
This work was partially completed during M. I.'s stay at the Joint 
Quantum Institute (National Institute of Standards and Technology and 
University of Maryland). J. K. F. acknowledges support under ARO 
Grant W911NF0710576 with funds from the DARPA OLE Program. Part of 
this work was completed during a stay at the Aspen Center for Physics.

\end{document}